# Transforming Commercial Contracts through Computable Contracting

John Cummins[1,2], Christopher D. Clack[2]

[1] Innovation Partners Ltd, London, UK
[2] Centre for Blockchain Technologies, Department of Computer Science, UCL, London, UK

E-mail: john@innovationpartners.co.uk; clack@cs.ucl.ac.uk

**Abstract**

Contracts are an essential and fundamental component of commerce and society, serving to clarify agreement between multiple parties. While digital technologies have helped to automate many activities associated with contracting, the contracts themselves continue, in the main, to be in the form of unstructured, natural-language text. This limits the scope for improvements in productivity and automation, as well as the emergence of new business models. To this end, this paper examines the concept of 'computable contracts' as objects that are understandable by both humans and computers, and goes on to present a framework that unifies a range of technologies and approaches that collectively will help to make computable contracting a reality.

## Introduction

Contracts are the bedrock of commerce: through helping to build a common understanding between multiple parties, contracts help to orchestrate the global exchange of tens of billions of dollars of goods and services around the world every single day (Statista). However, despite the ever increasing levels of digital connectivity, the drafting, reviewing, execution and monitoring of contracts remains relatively unchanged and has not kept pace with the levels of innovation seen in other areas of business.

While this paper is highly relevant to those directly involved in the management of contracts, the fundamental importance of contracts means that it is equally relevant to senior management and those charged with implementing digital transformation. In these professional contexts, this paper seeks to fulfil the following key objectives:

   i)   Develop a conceptual framework for computable contracting that unifies the many areas of relevant research and development in this space and to show how they are related.
   ii)  Based on this framework, identify some of the key considerations that will determine the success of adopting and implementing computable contracting approaches.

Figure 1 overleaf presents an innovation landscape for contracting technologies with two key dimensions: one that indicates the ease with which contracts can be understood and used by humans, and a second that indicates the degree to which contracts can be understood or processed by computers.

As a broad generalisation, the activity of contracting in the business world of today is characterised by the bottom-left quadrant. Created as monolithic, non-modular documents in natural language form, contracts are often difficult to read and to modify, and when they are modified, the process of doing so is clumsy, sometimes with unintended consequences such as inconsistencies and time delays. While 'vagueness' in contracts is sometimes a practical necessity, all too often it reflects the limited resources available for drafting, again with the possibility of costly dispute resolution actions further down the line. For the most part, contracts are isolated documents (either in hard or soft copy formats) with little connectivity to the business systems to which they refer.

The consequences of this innovation shortfall in contracting have been highlighted in a recent Harvard Business Review article on 'how AI is changing contracts' (Rich, 2018), which suggests that "inefficient contracting causes firms to lose between 5% and 40% of value depending on circumstances". Research undertaken by the International Association for Contract and Commercial Management (IACCM) (Cummins, 2012) concluded that: "good contract development and management could improve profitability by the equivalent of a massive 9% of annual revenue." Finally, according to KPMG research (KPMG, 2017): "ineffective governance of provider contracts can cause value leakage ranging from 17% to 40%."

Innovation in contracting has tended to fall into two main areas, either by moving horizontally or vertically from the bottom-left quadrant shown in Figure 1. *By moving horizontally and towards the right from this quadrant*, we are enabling humans to read and use contracts more easily. Typically, this might involve the creation of contracting standards, and indeed, there are many industry trade associations that take on this important role e.g. the International Swaps and Derivatives Association (ISDA, 2019) in financial services, or the Joint Contracts Tribunal (JCT, 2020) in the construction sector. Further innovations along the horizontal dimension might include the modularisation of contracts and the use of templating systems that can help to semi-automate the drafting of contracts. Contracts might also be annotated with metadata - information about the legal text. And, even the English used might be standardised in some way so as to avoid unintended ambiguities. Finally, and as the old adage suggests, 'a picture is worth a thousand words': the judicious use of carefully constructed graphics might actually make the act of reading contracts a more inviting experience, e.g. Lemonade contracts for insurance (Lemonade, 2020).

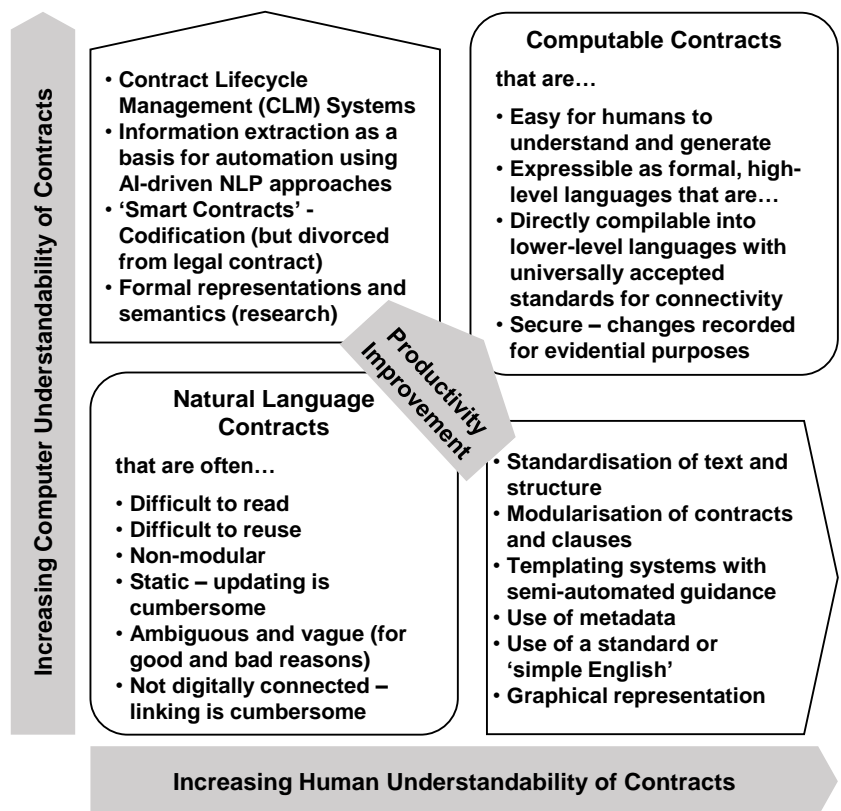

Figure 1: Innovation Landscape for Contracting Technologies

*A vertical transition upwards from the bottom-left quadrant* represents innovations that permit contracts to be understood by computers in some way. The most commonly deployed of these are contract lifecycle management (CLM) systems, essentially using document management approaches to underpin increased automation. Pushing forward on another front, and sometimes referred to as 'contract analytics', AI-based approaches are currently being deployed to help speed up the process of searching for, retrieving and analysing information from large bodies of text-based contracts (or 'large, unstructured datasets'). And finally, and perhaps the most fashionable at the moment, there are 'smart contracting' approaches, which involve the creation of pieces of computer code to enact specific aspects of a contract, and in some cases, mainly in fintech, are starting to muddy the distinction between computer code and legal contracts (Harley, 2017).

While these technology-driven solutions go some way towards addressing the innovation shortfall mentioned above, the scope for significant improvements in productivity is limited because of the way in which the vast majority of contracts are represented i.e. in a natural language form, with little, if any, formal structure or metadata. Hence, it is proposed that, only through fundamentally rethinking the way in which contracts are structured and expressed, can the full potential for productivity improvements in contracting be realised.

The *overall vision therefore is to reach the quadrant in the upper right*. This means the transformation of contracting from the creation of static documents, created in a natural language form and that often have limited direct linkage to business operations, to one that is driven by 'computable contracts' - readily understood by humans and computers, and that have a significantly greater level of digital connectivity. This transformative vision of computable contracting will facilitate productivity improvements and lower costs through reducing contract drafting times, increasing the scope for the automation of contract execution, and speeding up (or even avoiding) dispute resolution.

The term 'computable contracting' was first used and discussed extensively by Surden (2012). Here, Surden indicates that 'computable' means that the legal result is "***able to be automatically generated by a rules-based process***". And led in part by Goodenough (2019), the pioneering work of Surden continues to be advanced by the Codex Group at Stanford University.

Making contracts computable should not only bring about a profound and fundamental shift in the way that contracting in undertaken, but also pave the way for the emergence of new business models. In an analogous sense, it could do for contracting what computer aided design and production control systems have done for the automation of manufacturing.

## Review of the State of the Art

Innovation in the area of contracting in order to achieve productivity gains has generally pursued one or more of four main approaches from both product and business process perspectives: contract standardisation, contract lifecycle management, smart contracts and contract analytics.

**Contract standardisation** is essentially achieved through the creation of templates that can be used in a variety of similar situations. Either in hard-copy, or more often in digital formats, these templates (and templating systems) range in their levels of sophistication, but they all seek to facilitate the entry of information into a standardised 'data structure', such as a set of clauses with predefined fields. In its most basic form, it may be that the only information required for a template is the names of the contracting parties. More advanced, so-called 'contract assembly' systems, such as Hotdocs (2020), offer greater scope for contractual variation, often using modular ('lego-type') approaches for structuring contracts together with guidance on their use, that is both situation-specific and user-friendly. Using AI-driven technologies, Martin (2011-2012) has developed a comprehensive set of approaches for the deconstruction and standardisation of contracts, described more fully in the blog 'Contract Analysis and Contract Standards'.

**Contract Lifecycle Management (CLM)** systems often build upon contract templates to create a managed interface between a portfolio of contracts, on the one hand, and the relevant business systems on the other. At the heart of CLM systems is a set of managed data fields, from which data can be exported and imported, and therefore fed into (or drawn from) various automated or semi-automated business processes. In addition, some CLM providers offer relatively sophisticated contract assembly systems (as described above) as an initial step in managing the overall lifecycle of a contract. In their report on contract automation, the IACCM and Capgemini (2018, 1) suggest that there are over 200 vendors of CLM systems. This collaboration has led to the development of a 'contract automation software comparison tool' (IACCM Capgemini, 2018, 2).

**Smart Contracts** are concerned with helping to automate the performance (of part) of a contract and are perhaps the area most talked about in contract innovation. Although this term has been used for over two decades, coined initially by Szabo (1997) in the 1990's, the rapid increase in its current usage has been driven largely by the cryptocurrency community. Reflecting the work of Nick Szabo, the term 'smart contract' has been defined more generally by Clack et al (2016) to mean "an automatable and enforceable agreement. Automatable by computer, although some parts may require human input and control. Enforceable either by legal enforcement of rights and obligations or via tamper-proof execution of computer code."

In the cryptocurrency community, the term 'smart contract' has been used to refer to computer code that is sometimes (but not always) linked to a natural-language-based legal contract. In the work by Goldby et al (2019) in the insurance industry, smart contracts are defined as "pieces of computer code that are designed to start carrying out tasks automatically in response to external triggers". A good example here is perhaps the Accord Project (2020), with its "smart legal contracts stack" and templating system. Crucially however, for crypto-related smart contracts, the code is quite separate from the contract (or is an agreement that is not legally enforceable), and has resulted in several observers to suggest that "smart contracts are neither smart nor contracts" (Giancasproa, 2017).

From a legal perspective, there have been recent developments for smart contracts. In November 2019, the UK Jurisdiction taskforce concluded that smart contracts can be, or be part of, binding legal contracts. In the legal statement on cryptoassets and smart contracts (UK Lawtech Panel, 2019), the taskforce suggests that while "it is open to parties to agree expressly that a smart contract is not legally binding, as they could with a conventional agreement, it would be very unusual (and unwise) in a commercial context". In principle therefore, and subject to a range of points (that go beyond the scope of this paper), the ordinary rules of law apply to smart contracts.

**Contract analytics** concerns the use of a number of computational techniques that are used to establish insights into the content and structure of contracts. The broader term of 'AI' (artificial intelligence) for contracts usually concerns the application of machine learning to natural language processing (NLP). These AI techniques, coupled with substantial computing power, can be used for various forms of contract analysis, such as searching for specific information contained within large bodies of text. Martin (2011-2012) uses AI-based approaches to analyse contracts and to develop representative standards. Often, contract analytics is embedded within a broader CLM-based or workflow automation system, so that information extracted can then be used to help perform or trigger some kind of operation such as automated payments or email notifications.

### *State of the Art for Computable Contracts*

At a fundamental level, all automation is driven by a set of coded instructions. Hence, in order to automate the tasks specified in a contract, some form of machine-understandable code must be created, the functionality of which accurately reflects the content of a contract, or at least the automatable parts of it. In other words, it is vital that the set of coded instructions is an accurate translation of the contract (Farrell et al, 2019). This is not necessarily always the case, possibly because of ambiguity/vagueness in the contract, or poor coding, or misunderstandings between lawyers and programmers.

At the moment, and as reflected in this review of the state of the art, most of the approaches currently used to improve contracting efficiency are still built upon the fundamental notion that a contract is created and expressed using a natural language (such as English), both with its high level of expressivity (and concomitant scope for multiple interpretations), as well as its lack of direct machine-executability. *And crucially, any coding required to support automation is undertaken very separately from the creation of the contract.* There are, however, several exceptions to this general observation that will herald an evolution towards computable contracting, several key ones of which are mentioned in this section below.

The work by Surden has helped to pave the way for the Stanford Computable Contracts Initiative (SCCI, 2020), led in part by Goodenough. With a strong focus on use cases, the SCCI project is seeking to create contracts in a computable form from the outset. As suggested on the SCCI project page [22], "when terms and conditions are represented in highly-structured data, computers are able to process them automatically with a guaranteed degree of accuracy. The effect is not only a significant reduction of legal transaction costs, but it also opens a variety of new options to create better contracts".

Earlier work by Grigg (2000) in the late 1990's proposed the idea of a term 'Ricardian Contract' for financial trading. It is usefully defined as "a form of digital document that acts as an agreement between two parties on the terms and conditions for an interaction between the agreed parties." (Alam, 2018) It is also suggested that Ricardian Contracts are simultaneously understandable by humans and computers, which is achieved by adding tags and metadata to clauses (Lampič, 2019). Hazard has developed a library of 'prose objects' (Hazard, 2020) that have extensive tagging and exploit the parsing of natural language to facilitate computer readability.

In terms of languages that are simultaneously understandable by computers and humans, the very recent work by Diedrich (2020) on 'Lexon' digital contracts and Kowalski (2019) on the development of a form of 'logical English' are both noteworthy advances based on a substantial body of work. The domain-specific, contract specification language developed by Deon Digital (2020) in conjunction with the University of Copenhagen is a particularly good example of a language that straddles the divide between 'logical English' and standard computer code.

Other academic centres of excellence currently focussing on computable contracting approaches include the Financial Computing and Analytics Group in the Department of Computer Science at UCL (FCAG, 2020) in the UK and the work undertaken by CSIRO and the Data 61 initiative (2020) in Australia. There are also several other groups around the world working on important elements for computable contracting, particularly in the area of legal informatics, and many of whom are members of the International Association for Artificial Intelligence and Law (IAAIL, 2020) and have showcased their work at the International Conference on Artificial Intelligence and Law (ICAIL).

**Computable Contracts – A Vision**

As described in the top-right quadrant of Figure 1, computable contracts embody the following characteristics:
- Easy for humans to understand and generate;
- Expressible as a formal, high-level language that is directly compilable ('translatable') into lower-level languages;
- Adoption of universally accepted standards for connectivity (interoperability);
- Secure – changes recorded for evidential purposes.

Arguably, current approaches for DLT-based smart contracts can claim to have made significant progress in terms of interoperability and security. Crucially, however, it is the first two characteristics where computable contracting makes a significant and innovative step: ***contracts, expressed in a high-level, structured 'language', that can be a) readily generated, understood and used by humans, and b) easily and simultaneously understood (or processed) by computers.***

Making this vision a reality requires a somewhat broader view of contracting, one where contracts are seen as a fully integrated (or digitally connected) part of a commercial system and its associated business processes. And, as shown in Figure 2, this broader vision has four key components covering contracts, more specifically, and system-wide aspects, more generally, and all of which are subsequently discussed in turn.

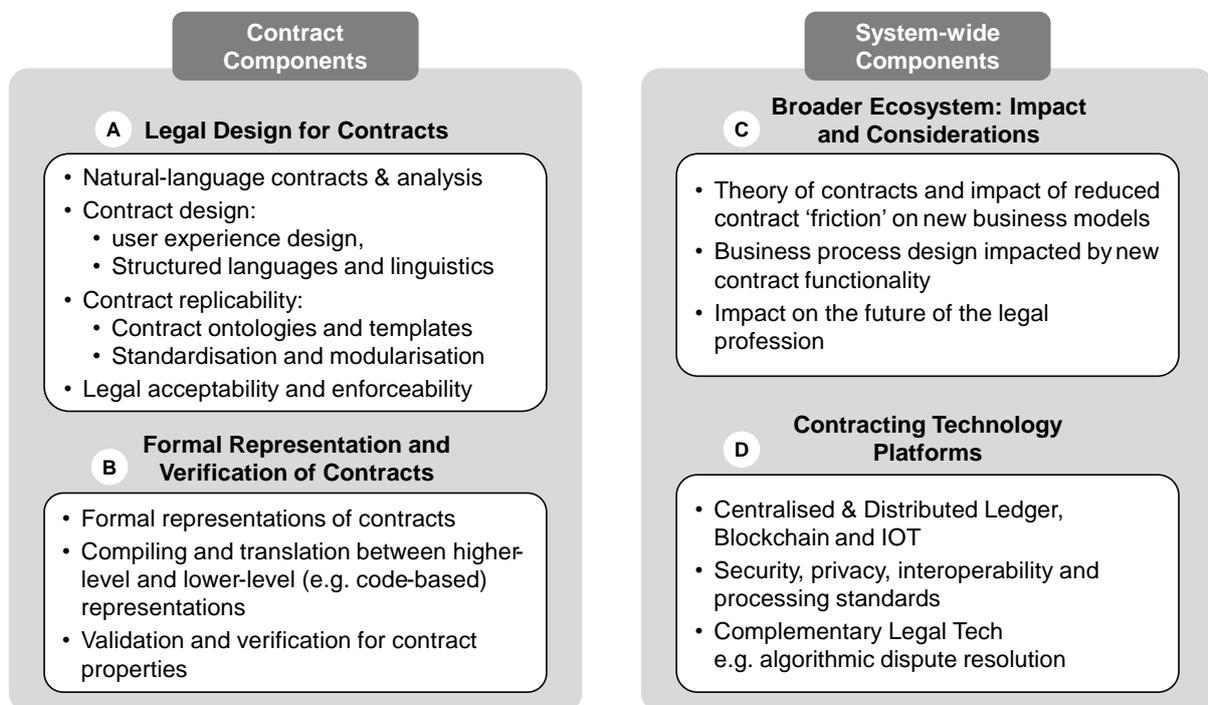

Figure 2: Vision for Computable Contracting

**A. Legal Design for Contracts**

This concerns all aspects that directly impact a lawyer, including both the study of traditional natural-language contracts and the study of contracts drafted in a more standardised manner. As indicated earlier, some of the AI-driven techniques for contracts analysis may continue to assist with contract standardisation. Importantly, computable contracting envisages that contracts will be partially or entirely expressed in some form of formal specification or high-level programming language, assisted with user-friendly interfacing, such as the pioneering work of Hagan and Haapio (2016). Furthermore, computable contracting does not mean that there is no room for vagueness, but more that vagueness, where it is deemed necessary, will be employed in a more deliberate way.

Potentially, a contracts development environment, somewhat analogous to what we currently see in the web development space, could emerge for legal professionals and will help to underpin adoption by a critical mass of practitioners. This is likely to require an intuitive, modular approach, supported by graphically-enhanced user interfaces as well as specially structured libraries of 'contract objects' validated by legal experts. Only when a critical mass of legal practitioners has been established will computable contracting achieve the required level of acceptability in the legal profession, which in turn will provide a solid platform for wider adoption.

### B. Formal Representation and Verification of Contracts

This covers the underlying science and logic concerning the structure and semantics of contracts, the development of formal analytical tools and formal verification and validation techniques. Developments in this area include translation from the high-level representations used in (A) to more formal representations (perhaps using formal logics or languages) and interaction with (A) in terms of reporting the results of analysis to lawyers. Work by Peyton-Jones et al (2000) focuses the application of functional programming techniques to financial contracts and has been an important catalyst in the emergence of domain-specific languages for defining contracts such as LexiFi's MLFi language (Eber, 2003)(Ali, Haldane, 2012). Of particular importance here is the work of Hvitved and Henglein (2014) at the University of Copenhagen in broadening the development of domain-specific languages for enterprise systems. From a more commercial perspective, the Zurich-based Deon Digital has developed these ideas into the CSL Language (Deon Digital, 2020); and in a similar vein, Digital Asset have developed their 'smart contracting' language, DAML (2020). Finally, a noteworthy example of a user-friendly and modular, contract specification language is Marlowe (Thompson, 2020).

More generally, the translation from higher-level representations (that are domain specific and user-friendly) to lower-level representations (in raw code) will provide more of a seamless link between a human understandable contract and automation across the broader ecosystem in areas such as negotiation, execution, contract performance and contract management.

### C. Broader Ecosystem: Impact and Considerations

This concerns research into issues of broader commercial economic and societal relevance, such as the way in which new contracting technology might alter the way in which firms do business and the size distribution of commercial entities. In 2016 The Nobel Memorial Prize in Economic Science was awarded to Oliver Hart and Bengt Holmström (Schmidt, 2017) for their work in building the foundations of contract theory, and serves as a reminder that, as technology improves and organisations become more complex, the theory and practice of contract design will only increase in importance.

This also includes implications for the legal profession and law. It has often been suggested that lawyers will have to learn how to code, or maybe that they should think in a structured way, as a programmer might. Arguably however, good lawyers (as do good programmers) already display a high level of structured thinking, and rather than 'getting lawyers to code', lawyers should be encouraged to think like designers. Indeed, the idea of 'Legal Design', championed by Haapio and Hagan (2016), has already had a significant impact on the legal community.

### D. Contracting Technology Platforms

The platform component comprises a range of facilitating technologies that support computable contracting such as distributed ledger technologies (DLTs); intelligent networks (including the 'Internet of Things' and the need for interoperability and security); and automated decision-making and machine-learning technologies.

Well-known platforms DLT platforms include Hyperledger, Corda, Cardano and Plutus. Corda builds on Grigg's work into Ricardian contracts (Grigg, 2000), as well as the work of Clack et al into smart contract templates (Clack et al, 2016). A deeper analysis of these technologies goes beyond the scope of this paper, but it is important to recognise that much of the current momentum in the area of smart or computable contracting stems from the growing excitement around blockchain and distributed ledger technologies (DLTs).

Academics from the Judge Business School have produced a conceptual framework (Rauchs, 2018) for DLTs and identifies five key characteristics that a DLT system must be capable of ensuring:

- Shared recordkeeping,
- Multi-party consensus on a shared set of records,
- Independent validation by each participant,
- Evidence of non-consensual changes, and
- Resistance to tampering.

While all of these characteristics are important for DLT, they are not necessarily a requirement for computable contracting. It is important to understand therefore, that while DLTs may help to accelerate the adoption of smart and computable contracting approaches, other technology platforms may also be appropriate.

*Vision - Technological Evolution towards Computable Contracts*

This evolution towards computable contracting, and the shift away from natural-language-based approaches that are typical in commerce today is shown in Figure 3. Moving from left to right, from 'business as usual', through various modes of 'smart contracting', and then onto 'unified contracts and code', the level of 'contract computability' increases.

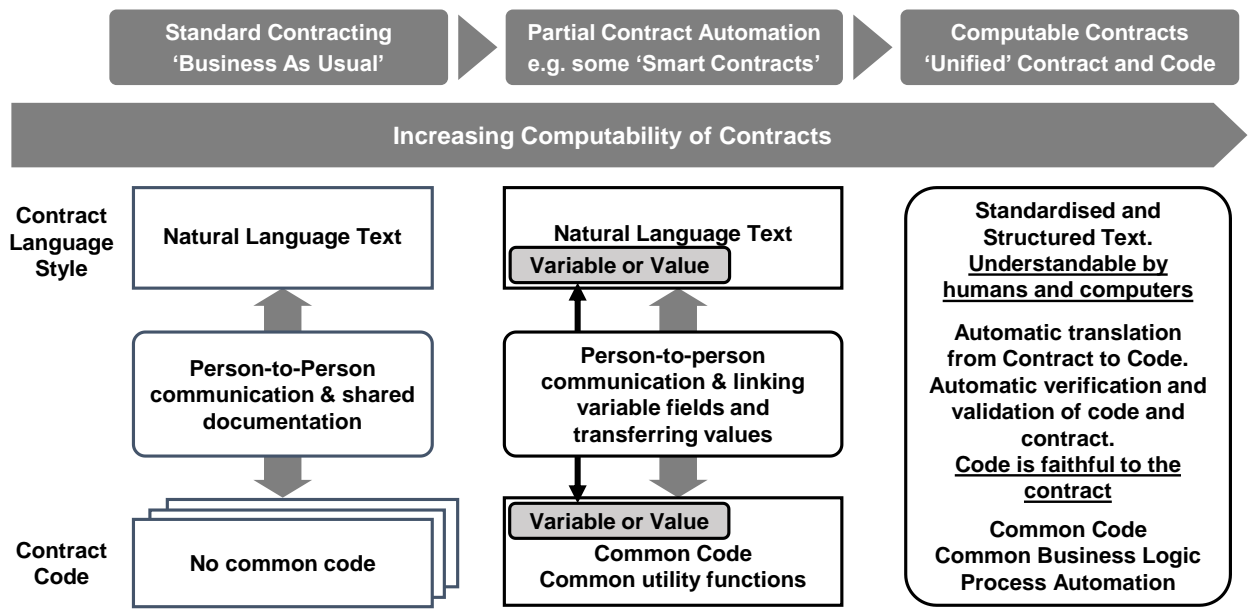

Figure 3: Technological evolution towards computable contracts

**Standard Contracting - 'Business as Usual'**

This scenario captures the bottom-left quadrant from Figure 1. It represents fundamental gaps in communication, not only between different organisational entities (where there is little or no common code), but also between lawyers and programmers. The scope of code use for contracting operations is typically limited, but where it is used, validation is time-consuming and expensive. With coding in different languages and multiple interpretations of contracts, there is a high likelihood of conflicts requiring costly and drawn-out human resolution.

**Partial Contract Automation - 'Smart' and Ricardian Contracting**

In this scenario, pragmatic approaches are taken to achieve 'partial automation', mainly through extracting information from defined information fields of contracts and feeding it onto wherever it is required in the broader system. Importantly, the establishment of common code across organisational boundaries has a significant and positive impact on overall efficiency. From a contracts perspective, the partial automation (e.g. through smart contracts) usually means identifying those parts of the contract where there is little complexity or vagueness (e.g. names or numerical values), and establishing various links to business processes, usually through some kind of tagging approach.

Current smart contracts fall short on two main fronts. The first problem is that only a relatively small part of the contractual information is captured for automation: current smart contracts tend to capture only a part of the operational aspects and rarely capture the deontic aspects concerning rights and obligations. This leads to a second problem concerning the piecemeal selection of operational information from defined fields in contracts: there is a tendency to ignore more complex (and perhaps less visible) conditionalities that are nonetheless related to the information that has been selected.

For some contracts however, and particularly financial contracts, where there is relatively little (or very simple) embedded logic concerning the operational information, Ricardian contracts using some form of 'tagging approach' (such as that proposed by Hazard) may well provide sufficient functionality for successful implementation (Alam, 2018) (Lampič, 2019) (Hazard, 2020).

**Computable Contracts – Unified Contracts and Code**

In the final scenario, the legal contract and the code are more unified with automated compiling (or translation) between the two. To enable this to happen, computable contracting approaches will tend to place a greater emphasis on the use of standardised or structured text that is more readily codified. Such approaches will include domain-specific languages and very high level programming languages that adopt a standard form of English, such as the 'Logical English' approach proposed by Kowalski (2019). Contract semantics will be captured formally, with flexibility relating to deliberately vague propositions. As these approaches are developed further, it is likely that a whole range of tried-and-tested, reusable 'legal objects' will be available, where the code is faithful to the contract, and which suit a whole range of deployment situations.

To get a sense of how the future may unfold for computable contracting, the core ideas behind the Semantic Web provide some important insights (Wikipedia). With the overall objective of enabling a computer to understand Internet data, the Semantic Web embodies a number of technologies that seek to build meaning and structure around this data. As indicated in the Wikipedia definition for 'semantic technology' (Wikipedia), "these technologies formally represent the meaning involved in information. For example, ontologies can describe concepts, relationships between things, and categories of things." Recent work by Casanovas and others (2016) explores the semantic web for the legal domain, while earlier work by Johannesson and Kabilan (2003), and more recently Griffo and others (2017), looks at modelling contracts using an ontology-based approach. Finally, in a presentation at the Internet of Agreements conference in 2017, Wong (2017) presents some useful ideas for a computable contracting technology stack, which shares several similarities with the Semantic Web technology stack.

**Computable Contracts – Key Considerations for Implementation**

As described in the It is hoped that the potential for significant improvements in operational efficiency and the opportunity to exploit new business models will encourage the implementation of computable contracting solutions in the foreseeable future (i.e. the next three to five years) across different industry sectors (e.g. finance, insurance, construction). This will include early-stage, prototyping deployments, as well as enhancements to existing systems.

However, before any initiative is undertaken to implement a computable contracting approaches, and as with any new business initiative, it is vital to determine whether or not it makes sense to do so.

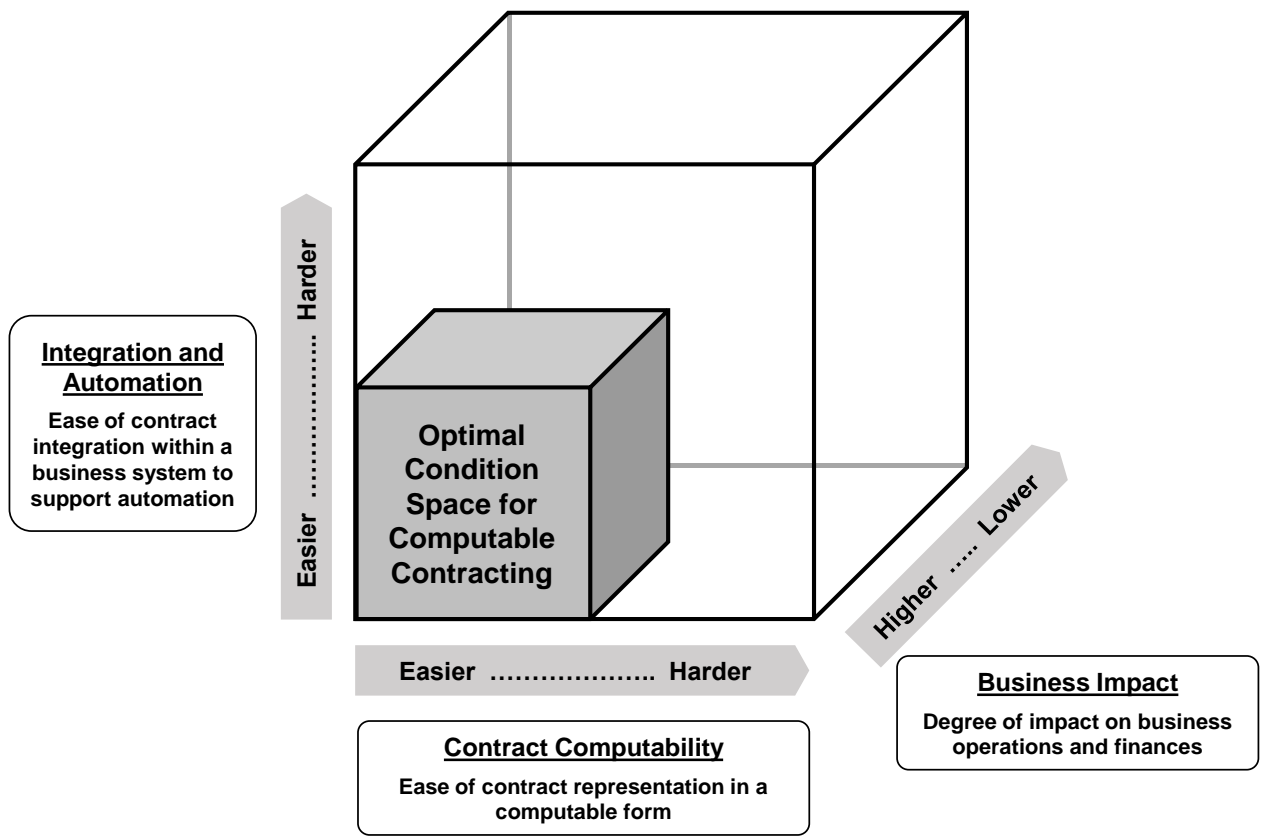

Figure 4: Key Considerations for Implementation of Computable Contracts

As a general guide, and as shown in Figure 4, there are three broad areas of consideration for the successful implementation of computable contracting, either for whole (or a suite of) contracts or for parts of a contract. The optimal conditions are found at the favourable ends of the spectrum for each of these three areas, namely:

**i. Contract Computability:** Ease with which a contract may be represented in a computable form, and where the representation of the contract can be understood simultaneously by computers and by humans. Important aspects to be assessed here are *consistency*, *completeness* (e.g. are all, or most, event states definable?) and *vagueness*.

**ii. Integration and Automation:** Ease with which a contract can be integrated with other components of a business system in order to support automation. Digital *connectivity* and *interoperability* are critical aspects to be addressed here.

**iii. Business Impact:** Degree of impact on business operations, which may be measured in terms of efficiency, productivity, quality and benefits; and financial performance, which may be measured in terms of cost savings and profitability.

Each of these three areas (contract computability; integration and automation; and business impact) are discussed in the next three subsections.

### *Contract Computability*

Making contracts computable means that the information contained within a contract can be understood by humans and computers. To achieve this objective, the following aspects should be considered:

- The main components of a contract;
- The consistency with terms are used (including vagueness and ambiguity);
- The complexity and completeness of a contract - covering both product and process complexity; and
- Domain-specific languages and user-friendly approaches for contract representation.

Each of these aspects is addressed in the following subsections.

### *Main Components of Contracts*

While the following is not an exhaustive list, it provides an overview of most of the components in a contract, and certainly those that are relevant for computability:

- **Contract objects** and their definitions - to clarify the meaning of key terms used in the contract e.g. an insurance contract may define specific types of insurance coverage.
- **Parties** - individuals or organisations between whom a contract exists.
- **Rights and obligations** of the parties (deontic aspects) - this deals with what is expected of the parties in order to fulfil the contract.
- **Operational aspects** - this covers *how* the contract obligations are to be fulfilled.
- **Temporal aspects** - this concerns aspects of time for fulfilment of the obligations e.g. what is actually meant by: 'within a seven day period'.
- **Events** - situations that may or should arise during the contract period.
- **Contract states** - may be thought of actual scenarios involving the parties, their fulfilment of obligations (or not), the operational and temporal aspects, and events. An example of a contract state might be: 'the supplier has not sent the shipment (obligation) with the required packaging (obligation/operational) and, due to bad weather (event), the shipment has arrived late (temporal)'.
- **Anticipated contract states** - potential scenarios that may arise and which are defined in the contract so that clarity is achieved on respective obligations in these specific scenarios.
- **Representations and warranties** concerning the above.

Representing the above information pertaining to a contract (or at least some of it) in a computable form has two main requisites: first, managing consistency, vagueness and ambiguity of contracts; and second, managing the complexity and completeness of contracts by defining a full or complete set of contract variations and states.

*Consistency, Vagueness and Ambiguity of Contracts*

While contracts may vary in their degree of complexity, there are also different ways of interpreting what 'complexity' of a contract actually means in practice. To start with, the perceived complexity of a contract may arise from inconsistent drafting, and where similar ideas are represented in different ways or with different terms. It may also be that, even though it displays internal consistency, a contract has terms that do not fall in line with industry standards (if indeed there are any!).

In order to represent contracts in a computable form, **consistency of expression is vital**. This is a first step and may be achieved without seeking any recourse to computational support. With reference to the Innovation Landscape for contracting technologies in Figure 1, this first step is represented as a horizontal move from the bottom left to the right.

In his work on contract deconstruction and analytics, Martin (2018) develops a two-dimensional visual analysis tool for the **consistency** (or standardisation) and **commonality** of specific provisions across a set of contracts. Here, it is worth noting that:
- The degree of consistency in a contract serves as an indication of the amount of preparatory work needed for making contracts computable; and
- The commonality of provisions, on the other hand, serves as more of a guide as to which parts of the contract might be prioritised for representation in a computable form.

Assuming that a high level of consistency has been established for a contract, the next thing to consider is vagueness. The ease with which contractual statements may be represented in a computable form *depends upon the extent to which they can be defined with sufficient specificity*. The term 'vague' is used to describe something that is not defined specifically e.g. 'both parties will take reasonable measures to ensure that the project progresses satisfactorily'. Both of the words 'reasonable' and 'satisfactorily' are vague, and often it may well be that the use of vague terms enables the parties to agree to the contract in the first place. Lawyers will also argue that, in specific situations, vague terms have specific meanings. To be pragmatic, therefore, a 'managed' approach should be taken, and where building in specificity makes sense to (and is agreeable by) all parties concerned.

With regard to managing vagueness, there are two approaches that can be pursued. **The first approach** is to define what is meant by 'reasonable' within the contract - internal management. This would require a set of clear statements on what the parties would actually consider 'reasonable' to mean in the context that it is used, and may involve the definition of multiple scenarios. **A second approach** is to define how disagreements that may arise will be managed, perhaps through both parties agreeing on some form of dispute resolution process or third party arbitration. This approach enables an objective view to be secured on whether or not something is reasonable, if and when such a view is required. This approach may also be fully or partially automated.

The term 'ambiguous' is used to describe something that, while appearing to be clear, is open to multiple interpretations e.g. 'the delivery should include A or B and C'. Whereas vagueness may require managing (or even be desirable), ambiguity should be avoided from the outset.

*Complexity of Contracts - Variations and States*

When a high level of consistency has been established for a portfolio of contracts (which in itself is no mean feat), a measure of complexity may be established through considering two further aspects: first, the **number of contract variations** (all of which must be defined consistently); and second, **the number of process states (or operational scenarios)** that are envisaged by the contract. This is shown in Figure 5, where the suitability of computable contracting for different modes of complexity is indicated.

**The lower, left-hand quadrant in Figure 5**

As a rule-of-thumb, it is easier to represent contracts with a lower level of complexity (i.e. contracts with minimal variations and contract states) in a computable form. However, for those contracts with a low level of complexity, there may be little added value of adopting computable contracting approaches beyond those contracting solutions currently deployed (and as shown in Figure 1).

For example, possibly the simplest form of contractual variation concerns the changing of one or more of the contracting parties across a suite of otherwise identical contracts. And it may also be that, irrespective of the contracting parties, the number and nature of operational scenarios is also identical in all cases. Here, given the success of many templating and smart-contract-based solutions, there is little justification for an alternative.

**Moving vertically or horizontally from the lower, left-hand quadrant in Figure 5**
More complex contracts are represented by moving horizontally to the right (increasing product complexity), or vertically upwards (increasing process complexity) in Figure 5. More complex contracts may also involve interdependencies e.g. between product variations and operational processes, or where a change in one aspect of a contract leads to several changes elsewhere.

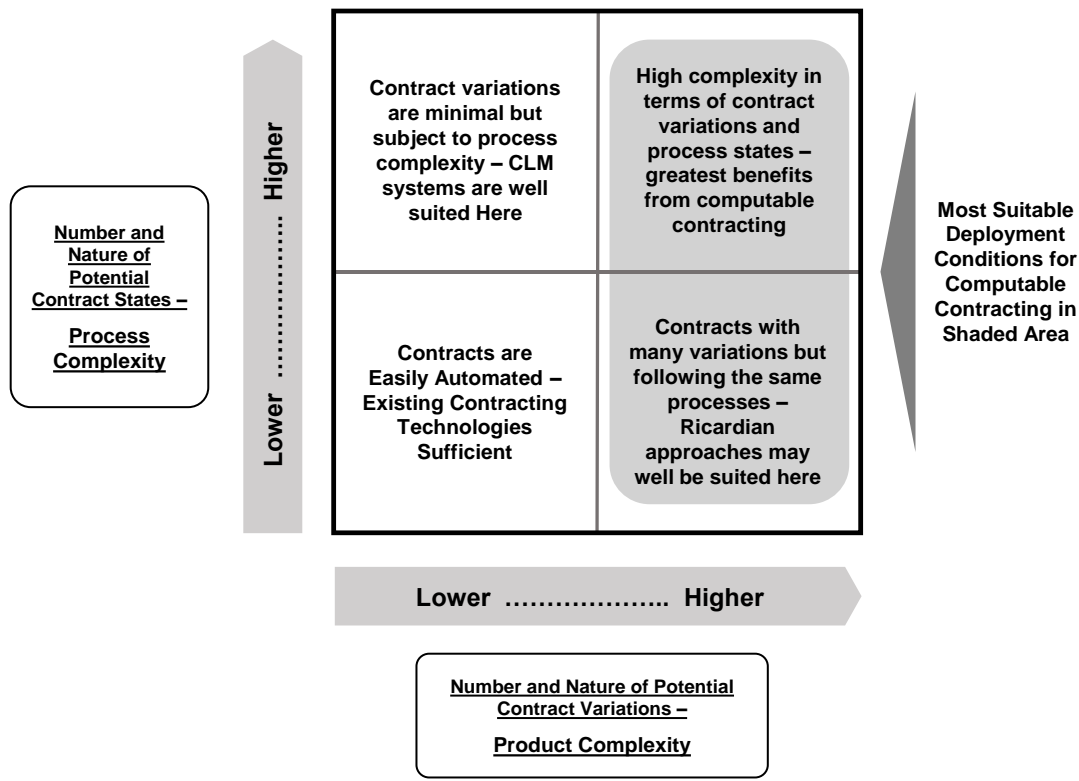

Figure 5: Contract Complexity and Computable Contracting

For deployment scenarios where the process complexity is high, but the product complexity is low, CLM systems are likely to be adequately suited. For the converse situation, where the process complexity is low, but the product complexity high, a Ricardian (or 'smart contracts') approach is likely to work well.

**'Complete' contracts and their representation**
With regard to the overall set of possible variations in a contract, the idea of 'completeness' is useful. In short, this means that all of the variations (either product or process related) of the contract can be specified (or anticipated) in some way. Furthermore, it may well be that there are thousands of potential contractual variations, but it is often the case that this is driven by a relatively small number (e.g. 10 or 20) of underpinning variables.

Various data structure formats may be used to fully represent the complete set of different contract variations and states. Contract variations may usefully be represented using a hierarchical approach. A tabular form may be used for contract-related process states, as well as those events which determine transitions between states, an example of which is given by Goodenough for loan contracts (2015). Here, the contract is viewed as an automaton and a full analysis of contract 'completeness' is undertaken by specifying all of the contract states and transition functions between the states.

In the previous subsection, the importance of standardisation or consistency was emphasised. Indeed, without the shared adoption of standards, there can be no implementable form of computable contracting, whatever the level of complexity of the contracts. A good example of the development of standards to underpin digital transactions is the work of ISDA (International Swaps and Derivatives Association) to establish a 'Common Domain Model (CDM)'. The CDM will provide "a single, common digital representation of derivatives trade events and actions, and will enhance consistency and facilitate interoperability across firms and platforms" (ISDA, 2020). Hence, events and processes throughout the lifecycle of the trade of a derivative or swap are standardised under the ISDA CDM and operate on the same shared data, or on data that is consistent with the ISDA CDM format.

*Domain-Specific Languages and User Friendliness*

As indicated earlier in this document, most of the existing computer code relating to contracts and their automation is wholly separate from the contract with few, if any, links. This raises the fundamental question as to whether or not the contract has faithfully represented by the commercial system(s) and any operations undertaken. It has already been emphasised that computable contracting aims to integrate both the contract and the code in a more unified, high-level representation that is understandable by both humans and computers. One way of achieving this is through the use of a standard (or domain-specific) language (rather than an unconstrained natural language) in order to faithfully represent the different components of a contract (as outlined in the earlier subsection).

In his work on contract formalisation, Hvitved (2012) looks at the suitability of formal languages (and models) for the representation of various aspects of contracts. His work demonstrates that the type of formal language used (and therefore the nature of the domain-specific language) depends upon the required functionality of the contract. In other words, a domain-specific language is not suited because it has the right vocabulary, but more because it has the right grammar (i.e. how the language works at a more fundamental level). The work by Kowalski (2019) to develop a standard (or 'logical') form of English is relevant here.

In order to facilitate the re-use of elements of a computable contract and the establishment of shared/universal standards, a domain-specific language should facilitate 'modular approaches'. Hvitved also addresses the important aspect of modularity of contracts in his PhD Thesis (2012), and the concept is also explored by Smith (2006), who suggests that modularity "plays a pervasive role in managing contractual complexity".

Modularity also concerns the creation and re-use of larger contract 'blocks' that comprise smaller objects or modules, each of which fits together naturally to provide a functionality that is greater than any provided by the individual objects/modules. Furthermore, the creation of these larger blocks might be made easier through the use of 'wizards' (expert systems employing rules-based and/or probabilistic approaches) to assist the contract designer. Such approaches are already used to some extent in contract drafting.

A good example of a domain-specific language with a high-level language representation, and that adopts a modular, 'building-block' (almost literally by using Blockley) is Marlowe (Thompson, 2020), developed by Simon Thompson from the University of Kent in the UK. Marlowe is used for modelling financial instruments as smart contracts, and importantly has "been designed for people who are business engineers or subject experts rather than experienced developers", with little, if any, need for programming experience. With regard to modularity, Marlowe "comprises a small number of powerful building blocks that can be assembled into expressive financial contracts". Another example of a domain-specific language in its early stages of development is L4 from Legalese (Wong, 2020). Their ambition is undoubtedly to go beyond smart and Ricardian contracting and to change the way in which legal documents are expressed from the outset.

Continuing with the example of establishing standards in the previous section, ISDA have also established a platform (ISDA Create) (ISDA, 2020) to produce, deliver, negotiate and execute derivatives documents in a digital environment. The use of a common platform provides for increased efficiencies in both transactional and decision-making activities.

Graphical approaches may also be used to extend the expressivity of any standard language and will contribute towards the definition and validity-checking of both the variables and the logic of a contract. While it is not common practice at the moment, the use of innovative graphical approaches to enhance clarity and computability would be welcomed by the contracting parties and contract designers alike. For example, Wong, Haapio and others have worked collaboratively (Wong et al, 2015) to bring together concepts from contract visualisation and computable contracting.

*Integration and Automation for Computable Contracting*

The second key area to be considered in implementing computable contracting concerns integration and automation, each of which is dealt with in turn.

*Integration through Digital Connectivity*

For a computable contract to be useful, it ***must*** have digital connectivity. In other words, even though a contract may have been represented in a form that could (in theory) be understood by a computer, it will still essentially be a 'dumb' object if it cannot be connected to the system(s) to which it refers (e.g. a payments system, or a delivery registration system). Hence, the ease with which this connectivity can be established (and thereby integrating the contract with other components of a business system in order to support automation) will determine whether or not a contract can be made computable.

One of the chief determinants of connectivity is interoperability i.e. whether the individual components (including the computable contract itself) can successfully communicate with one another. This need for interoperability will also be driven by the emergence of what is sometimes referred to as the 'internet of things', a global network of devices and systems, each with its own unique IP address. Arguably, this fundamental shift towards a more connected system of devices across a network may be pivotal in heralding a new paradigm in contracting.

*Automation*

With the increased use of automatic devices and controls in mechanised production lines in the automobile industry, the term 'automation' became commonly used in the late 1940s. Going back further still, the word 'automate' comes from the Greek word 'automatos', which means 'to act of itself' or demonstrating self-governance. Automation is now used to describe the application of machines to tasks once performed by human beings or, increasingly, to tasks that would otherwise be impossible for humans to undertake. And, driven by the two key drivers of quality and efficiency improvements, there is scarcely an area of modern life that has not been affected by automation.

Curiously, and until fairly recently, the practice of law, had managed to resist any significant changes driven by automation, apart from maybe for the use of fairly standard, document management, word processing and electronic mailing systems. Over the last decade, however, this has started to change; law firms and paralegal teams are now fully embracing the world of 'legal tech'. This has already had a significant impact on case management through the use of intelligent and automated tools for the search and retrieval of information. And, as discussed earlier, contract analytics is now common practice, as is automated dispute resolution.

As with any process, the automation of contracting processes requires a thorough understanding of how activities fit together, the inputs and outputs, and how these might vary as circumstances change. For example, this requires all event states referred to in the contract to be defined, and for events to be monitored, categorised and registered. However, where we cannot anticipate every future eventuality (or do not wish to spend the time doing so) a managed approach towards dealing with vagueness will be needed.

Even when a contracting system is operating within the bounds of a defined set of states, there may well be a need for human intervention, particularly where decisions with significant consequences e.g. sizeable transfers of funds or assets are necessary, or where additional human judgement is required. Human intervention can be designed into processes from the outset, and follows an increasing trend of collaboration between computers and humans (Deloitte, 2019). For example, the term 'cobot' is used to describe a robot that can collaborate with a human.

Successful deployments of computable contracting may also consider compliance as part of the overall system functionality. In the same way that contracts can be represented in a computable way, so too can some of the legislation to which contracts should adhere. Indeed there is a significant level of interest in the emerging area of 'reg tech', where the computability of regulations (rather than contracts) is very much centre stage. Hence, forward-thinking deployments of computable contracting system should build in the connectivity to (and interoperability with) guiding legislation. One of the world leaders is this space is Csiro and the Data 61 Initiative (2020).

Automated contracting systems need to be trusted, and it is here that distributed ledger technologies (DLT) have a role to play. Having an immutable record of the dynamic states of a system will help to establish the necessary levels of trust so that multiple parties are willing to use it. There are several different DLT types, each suited to specific implementation scenarios (Governatori, 2018), and understanding how to design and build a system with the required levels of trust and security for different contracting scenarios between multiple parties is vital.

*Business Impact of Computable Contracting*

It may well be that a set of contracts lends itself to computability, both in terms of its formal representation and its digital connectivity, but there is one further hurdle to clear: does it make business sense? Without a high level of impact on business operations (in terms of efficiency, productivity, quality and benefits) and financial performance (costs and profits), any attempt to make contracts computable is likely to fail.

Hence, computable contracting initiatives need an appropriate costs/benefits analysis that extend across the full value chain, rather than considering improvements in, for example, the drafting or amendment of contracts alone. One aspect to be considered here may concern any reductions in contractual disputes (and the resources associated with them).

From a senior management perspective, computable contracting will provide a better platform for decision making. The data models and structures that are embedded in computable contracts can be readily aggregated and expressed in a dashboard format to allow for a more comprehensive coverage of the decision-making space, greater accuracy and increased responsiveness.

Computable contracting will also lead to the development of new business models, and while difficult to predict, let alone quantify, it is here perhaps that the most significant benefits will be found. A good example can be found in the insurance sector where there is significant latent potential for an increased level of capitalisation options for some types of more complex risk. One essential element in achieving this requires the underlying contracts to be built upon more standardised provisions and a shared set of data structures (Lloyds of London, 2019).

Achieving this business impact also requires organisational change, but there may well be resistance towards adopting some of the approaches necessary for computable contracting. Without a strong drive to establish standards (from the top), as well as a shared incentive to adopt these standards (throughout the organisations concerned), then change will be slow and piecemeal. And for many industries, even though a strong drive towards establishing standards from the top may well exist, it is the shared incentives (or ensuring benefits to all) that is usually the stumbling block. It follows therefore, that successful implementation of computable contracting solutions will require a combination of ambition, realism and pragmatism.

**Summary and Conclusion**

In the last three decades or so, digital innovations of all forms have had a transformational impact on our lives and the way in which commerce is undertaken. As a cornerstone of business, commercial contracts, however, have been left somewhat behind: they are generally created in a monolithic (rather than a modular) form and have little inbuilt structure and intelligence that might directly underpin greater digital automation.

Where advanced technologies have been used (e.g. contract lifecycle management systems), they have tended to automate around the contracts, leaving the contracts themselves largely untouched in their unstructured, natural-language form.

This shortfall in digital innovation for contracting is one of the causes of inefficiency and loss of productivity in commerce, and it is also holding back the emergence of more agile business models. Various studies over the past five to ten years have concluded that inefficient contracting causes double-digit losses in a wide range of industries.

Computable contracting sets out to close this productivity gap by transforming contracts into digital objects that are:
  i) Easy for humans to understand and generate;
  ii) Expressible (as far as possible) as a formal, high-level language that is directly compilable ('translatable') into lower-level languages.

To do this will require the adoption of universally accepted standards for connectivity (and interoperability) throughout the relevant community of interest, as well as providing the required levels of security.

Designing (or 'drafting') computable contracts is likely to borrow some of the ideas from the Semantic Web movement, with its associated technology stack (higher to lower level coding languages), and rich set of functionalities (such as ontology development and cryptography). Natural language may still be used in contracts, if only to provide annotated descriptions for the contract, but many aspects of contracts will have greater structure and shared formats from the outset. Hence, rather than dispensing with lawyers, the future of (computable) contracting will call upon new skills; the creativity in drafting contracts will be deployed in designing multi-facetted digital objects that are fully interoperable across business systems and the value chain.

The evolution towards computable contracting will take years, and will have to contend with resistance to change from several quarters, and not only the lawyers. It will build on many of the contracting technologies already being used today such as contract lifecycle management systems, smart and Ricardian contracting approaches, and automated document assembly systems. It will draw upon many of the ideas and practices commonly seen in the web development space, as well as the emergence of new coding languages. And, it will emerge more quickly in some industries than it will in others: financial services is likely to lead the way.

Building structure is about taming complexity, and for contracting, it is no different. On the assumption that standards can be established, the task of making contracts computable must deal with those aspects of a contract that concern the 'product' (i.e. what the obligations are) and the 'process' (i.e. how the obligations are met). Computable contracting will have the biggest impact initially where the product complexity is higher and the process complexity lower, and this is where much of the innovation in fin-tech and insure-tech is seen.

Future directions in research for computable contracting will be strongly focussed on the representation of the knowledge contained in the contract and related documentation, and in particular, the selection of the most appropriate representational form to suit the purpose. The three main forms to choose from are:

   i.   Controlled natural language – essentially, a natural language in a structured format,
   ii.  Natural language with embedded metadata, and
   iii. Graphical representations.

Building contracts that bring together all of the above in the most appropriate way will herald the emergence of a new skill of contract engineering and design.